\documentclass[12pt,a4paper]{article}

\usepackage{a4wide,cite}
\usepackage{epsfig}
\newcommand{\Li}[2]{{\mbox{Li}}_{#1}\left(#2\right)}
\newcommand{\Cl}[2]{{\mbox{Cl}}_{#1}\left(#2\right)}

\newcommand{\be}{\begin{equation}}
\newcommand{\ee}{\end{equation}}
\newcommand{\bea}{\begin{eqnarray}}
\newcommand{\eea}{\end{eqnarray}}
\newcommand{\ep}{\varepsilon}

\newcommand{\nn}{\nonumber}

\begin{document}

\renewcommand{\thefootnote}{\fnsymbol{footnote}}                                
 \thispagestyle{empty}
 \begin{flushright}
 {MZ-TH/01-14} \\[3mm]
 {hep-ph/0105072} \\[3mm]
 {May 2001}
 \end{flushright}
 \vspace*{2.0cm}
 \begin{center}
 {\large \bf
Quark mass dependence of the one-loop three-gluon vertex \\[3mm]
in arbitrary dimension}
 \end{center}
 \vspace{1cm}  
 \begin{center}
 A.I.~Davydychev$^{a,}$\footnote{On leave from
 Institute for Nuclear Physics, Moscow State University,
 119899, Moscow, Russia. Email address:
 davyd@thep.physik.uni-mainz.de}, \ \
 P.~Osland$^{b,}$\footnote{Email address: Per.Osland@fi.uib.no} \ \
 \ \ and \ \
 L.~Saks$^{b,}$\footnote{Email address: Leo.Saks@fi.uib.no}
\\
 \vspace{1cm}
$^{a}${\em
 Department of Physics,
 University of Mainz, \\
 Staudingerweg 7,
 D-55099 Mainz, Germany}
\\
\vspace{.3cm}  
$^{b}${\em
 Department of Physics,
 University of Bergen, \\
 All\'egaten 55,
 N-5007 Bergen, Norway}
\\
\end{center}
\hspace{3in}
\begin{abstract}
The one-loop off-shell massive quark contribution to the
three-gluon vertex is calculated in an arbitrary space-time 
dimension. The results for
all relevant on-shell and symmetric limits are obtained
directly from the general off-shell results. 
The analytic structure of the results for the relevant
massive scalar integrals is also discussed.
\end{abstract}
 
\newpage
\renewcommand{\thefootnote}{\arabic{footnote}}
\setcounter{footnote}{0}
\section{Introduction}

The three-gluon vertex is the basic object responsible for 
the non-Abelian nature of Quantum Chromodynamics~\cite{QCD}. 
Perturbative corrections to gluonic vertices are very important 
in real physical calculations, 
such as multijet production at the hadron colliders
(see e.g.\ \cite{QCD-reviews} and references therein). 
At the present high level of accuracy, one needs to perform not only 
calculations with on-shell external particles, 
there are also contributions where general off-shell results
are needed.

Several special cases for the one-loop three-gluon vertex have been
known already for a couple of decades.
Previous studies of the
three-gluon vertex have mainly been carried out with massless quarks, 
or with no quarks at all\footnote{Here and below,   
we mainly discuss the results in covariant gauges.}.  
Around 1980, Celmaster, Gonsalves, Pascual and Tarrach studied
the one-loop three-gluon vertex with massless quarks 
in the symmetric limit, $p_1^2=p_2^2=p_3^2$, mainly for the purpose
of comparing different renormalization schemes \cite{CelGon,PT}. 
The results for the case when two gluons are on shell have been 
given by Nowak,
Prasza{\l}owicz and S{\l}omi{\'n}ski \cite{NPS}.  In the pure 
gluodynamics 
(leaving out the quark loops), Ball and Chiu considered the
off-shell case in the Feynman gauge \cite{BC2}. Brandt and Frenkel
presented results for the infrared-singular parts with one and two
on-shell gluons in an arbitrary covariant gauge \cite{BF}.

More recently, general one-loop results for the three-gluon
vertex, in an arbitrary covariant gauge and dimension, have been 
presented 
in~\cite{DOT1} (see also \cite{DOT-Rh} for a brief review). 
Table~1 of~\cite{DOT1} gives an overview of the results 
for the one-loop three-gluon vertex obtained in the preceding
papers. The results of~\cite{DOT1} have contributed 
towards covering all remaining ``white spots'' in Table~1,
with only one restriction:
the case of massless quarks was considered.  

The purpose of the present paper is to extend the work started 
in~\cite{DOT1} and complete the study of the one-loop-order 
three-gluon vertex, by considering massive
quarks in the quark-loop contributions,  
for an arbitrary value of the
space-time dimension. We note that these contributions do not
depend on the gauge parameter at one loop. 
Thereby, we would allow for non-zero quark 
masses in all configurations listed in Table~1 of \cite{DOT1},
including the most general off-shell case.
Taking into account that the corresponding investigation 
of the quark-gluon vertex with massive quarks is given in~\cite{DOS},
and remembering that the one-loop ghost-gluon vertex~\cite{DOT1}
does not
involve any quark contributions, one can see that this is
indeed the last step, in addition to Refs.~\cite{DOT1,DOS}, needed 
to complete the most general study of the one-loop three-point 
vertices in QCD.

Such general results for the one-loop
vertices can be useful in the evaluation of two-loop (or higher)
order corrections. In particular, they can be used as ``blocks''
in evaluating higher-order corrections in QCD.
Presenting results in arbitrary dimension $n$,
we can obtain further terms of the expansion in $\ep=(4-n)/2$,
in the framework of dimensional regularization~\cite{dimreg}.  
Moreover, we can derive results for all on-shell limits 
of interest directly from the general ones. From arbitrary dimension
one can go over to two and three dimensions which are also
investigated in the context of QCD (see, e.g., in~\cite{AA}).
Finally, we would like to mention recent progress in 
the lattice calculations related to the QCD vertices
(see, e.g., in Refs.~\cite{lattice}).

We also discuss the analytic structure of the results for the
massive scalar integrals which appear in the calculations, 
using the geometrical approach of 
Ref.~\cite{DD}. Worth noting is that
scalar integrals of this type occur not only in the  
three-gluon coupling but also in one-loop corrections to
$H\gamma\gamma$, $HZ\gamma$,
$ZZH$, $W^{+}W^{-}H$ and some other 
vertices (see, e.g., in~\cite{Kn,HHG}).

The rest of the paper is organized as follows.  In Section~2, we give 
the notation for the three-gluon vertex, present the relevant
Ward--Slavnov--Taylor identity, and describe the tensor decomposition.
In Section~3, we list the most general off-shell results, and the
corresponding on-shell limits for the vertex.  The scalar integrals
are discussed in Section~4.  Conclusions and a summary are given in
Section~5.

\section{Preliminaries}
\setcounter{equation}{0}

\subsection{Decomposition of the three-gluon vertex}

Relevant notation for the three-gluon vertex, 
\bea
\label{ggg_general-Gamma}
{\Gamma}^{a_1 a_2 a_3}_{\mu_1 \mu_2 \mu_3}(p_1,p_2,p_3)
\equiv - \mbox{i} g f^{a_1 a_2 a_3} 
{\Gamma}_{\mu_1 \mu_2 \mu_3}(p_1,p_2,p_3),
\eea
is given in Fig.~1. 
We note that all gluon momenta are ingoing, $p_1+p_2+p_3=0$.
There are actually two diagrams, the fermion lines may be oriented
either way. Because of the color factors, Furry's theorem does
not apply, and the two diagrams add to produce a factor of two.
\begin{figure}[htb]
\refstepcounter{figure}
\label{Fig:ggg}
\addtocounter{figure}{-1}
\begin{center}
\setlength{\unitlength}{1cm}
\begin{picture}(10.0,5.5)
\put(1,-0.5){
\mbox{\epsfysize=7cm\epsffile{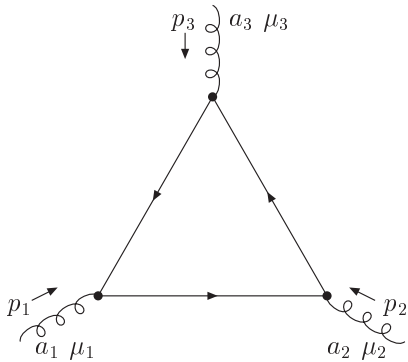}}}
\end{picture}
\vspace*{-4mm}
\caption{Notations used for the three-gluon vertex.}
\end{center}
\end{figure}

For the off-shell three-gluon vertex,
we adopt the well-known decomposition 
proposed by Ball and Chiu~\cite{BC2}\footnote{Another
general decomposition of the three-gluon vertex was considered
in Ref.~\cite{KB}.},
\begin{eqnarray}
\label{BC-ggg}
\Gamma_{\mu_1 \mu_2 \mu_3}(p_1, p_2, p_3)
= A(p_1^2, p_2^2; p_3^2)\; g_{\mu_1 \mu_2} (p_1-p_2)_{\mu_3}
+ B(p_1^2, p_2^2; p_3^2)\; g_{\mu_1 \mu_2} (p_1+p_2)_{\mu_3}
\hspace{10mm}   
\nonumber \\
- C(p_1^2, p_2^2; p_3^2)
\left[ (p_1 p_2) g_{\mu_1 \mu_2} - {p_1}_{\mu_2} {p_2}_{\mu_1} \right]
(p_1-p_2)_{\mu_3}
\hspace{57mm}
\nonumber \\
+ \textstyle{1\over3}\; S(p_1^2, p_2^2, p_3^2)
\left( {p_1}_{\mu_3} {p_2}_{\mu_1} {p_3}_{\mu_2}
     + {p_1}_{\mu_2} {p_2}_{\mu_3} {p_3}_{\mu_1} \right)
\hspace{64mm} 
\nonumber \\
+ F(p_1^2, p_2^2; p_3^2)
\left[ (p_1 p_2) g_{\mu_1 \mu_2} - {p_1}_{\mu_2} {p_2}_{\mu_1} \right]
\left[ {p_1}_{\mu_3} (p_2 p_3) - {p_2}_{\mu_3} (p_1 p_3) \right]
\hspace{32mm}
\nonumber \\
+ H(p_1^2, p_2^2, p_3^2)
\left\{ -g_{\mu_1 \mu_2}
\left[ {p_1}_{\mu_3} (p_2 p_3) \!-\! {p_2}_{\mu_3} (p_1 p_3) \right]
+ \textstyle{1\over3}
\left( {p_1}_{\mu_3} {p_2}_{\mu_1} {p_3}_{\mu_2}
\!-\! {p_1}_{\mu_2} {p_2}_{\mu_3} {p_3}_{\mu_1} \right) \right\}
\nonumber \\
+ \left\{ \; \mbox{cyclic permutations of} \; (p_1,\mu_1),
(p_2,\mu_2), (p_3,\mu_3)\; \right\}  .
\hspace{20mm} 
\end{eqnarray}
Here, the $A$, $C$ and $F$ functions are symmetric in the first two
arguments, the $H$ function is totally symmetric, the $B$ function
is antisymmetric
in the first two arguments, while the $S$ function is antisymmetric
with respect to interchange of any pair of arguments\footnote{In
Ref.~\cite{DOT1} it was shown that the one-loop $S$ function
vanishes at the one-loop order. This is also the case for
massive quarks (see below).}.
Note that the contributions containing the $F$ and $H$ functions
are totally transverse, i.e., they give zero when contracted with
any of ${p_1}_{\mu_1}$, ${p_2}_{\mu_2}$ or ${p_3}_{\mu_3}$.

\subsection{Basic integrals}

For the basic integrals, we follow the notation of Refs.~\cite{DOT1,DOS}
(see also in~\cite{BD-TMF}).
The scalar three-point integrals with equal masses associated with
all three internal lines are defined as
\bea
\label{defJ3}
J_{3}(\nu_1,\nu_2,\nu_3)
\equiv 
\int 
\frac{\mbox{d}^n q}{\left[(p_2-q)^2-m^2 \right]^{\nu_1}
\left[(p_1+q)^2-m^2 \right]^{\nu_2}
[q^2-m^2  ]^{\nu_3}}.
\eea
Here and henceforth, the causal prescription is understood,
$1/p^2 \leftrightarrow 1/(p^2 +{\rm i}0)$.

The following massive integrals are involved in the three-gluon vertex
calculation:
\bea
\label{J_3 defs}
J_3(1,1,1)&=&\mbox{i}\; \pi^{n/2} \;  \eta \;  
\varphi_3 \; ,
\nn \\	   
J_3(0,1,1)&=&\mbox{i}\; \pi^{n/2} \;  \eta \;  \kappa_{2,1} ,
\nn \\ 	   
J_3(1,0,1)&=&\mbox{i}\; \pi^{n/2} \;  \eta \;  \kappa_{2,2} ,
\nn \\   
J_3(1,1,0)&=&\mbox{i}\; \pi^{n/2} \;  \eta \;  \kappa_{2,3} ,
\nn \\ 	   
J_3(0,0,1)&=&\mbox{i}\; \pi^{n/2} \;  \eta \;  m^2 \; \widetilde{\kappa} ,
\eea
where the overall factor $\eta$ (depending on $n=4-2\ep$) is defined by
\begin{equation}
\label{eta}
\eta \equiv
\frac{\Gamma^2(\frac{n}{2}-1)}{\Gamma(n-3)} \;
     \Gamma(3-{\textstyle{n\over2}}) =
\frac{\Gamma^2(1-\varepsilon)}{\Gamma(1-2\varepsilon)} \;
\Gamma(1+\varepsilon) .
\end{equation} 
In Eqs.~(\ref{J_3 defs}),
$\varphi_3\equiv\varphi_3(p_1^2,p_2^2,p_3^2;m)$ 
is the non-trivial function
associated with the scalar triangle integral. The subscript ``3'' 
indicates that all three internal lines are massive.
This three-point function is discussed in more detail in Sec.~4.
For the two-point integrals we introduce the functions
\begin{equation}
\kappa_2(p_l^2;m)\equiv \kappa_{2,l} ,
\end{equation}
where $p_l$ ($l=1,2,3$) is the external momentum of the two-point
function, whereas the subscript ``2'' shows that both
internal propagators are massive.
The relevant information about this two-point function
can be found in Appendix~C1 of Ref.~\cite{DOS}.
Finally, $\widetilde{\kappa}$ corresponds to the tadpole 
contribution,
\be
\label{kt def}
\widetilde{\kappa} \equiv \widetilde{\kappa}(m^2)
\equiv \frac{\Gamma(1-2\ep)}{\Gamma^2(1-\ep)}\;
\frac{1}{\ep(1-\ep)}\; (m^2)^{-\ep} .
\ee
As a rule, we will omit the mass arguments in $\varphi_3$, $\kappa_2$
and $\widetilde{\kappa}$.  

\subsection{Ward--Slavnov--Taylor identity}

The Ward--Slavnov--Taylor (WST) identity for the three-gluon
vertex reads (see, e.g., in Refs.~\cite{MarPag,BC2,PT-QCD}) 
\bea
\label{ggg_WST}
p_3^{\mu_3} \Gamma_{\mu_1 \mu_2 \mu_3}(p_1, p_2, p_3)
&=& 
- 
\left( g_{\mu_1 }^{ \mu_3} p_1^2
- {p_1}_{\mu_1}  p_1^{\mu_3} \right) 
J(p_1^2)  G(p_3^2) 
\widetilde{\Gamma}_{\mu_3 \mu_2}(p_1, p_3; p_2)
\nn \\
&& 
+ 
\left( g_{\mu_2 }^{ \mu_3} p_2^2
       - {p_2}_{\mu_2}  p_2^{\mu_3} \right) 
J(p_2^2)  G(p_3^2) 
\widetilde{\Gamma}_{\mu_3 \mu_1}(p_2, p_3; p_1) ,
\eea
where $J(p^2)$ and $G(p^2)$ are scalar functions 
in the gluon polarization operator and the ghost self-energy,
respectively:
\bea
\label{gluon self-energy}
\Pi_{\mu_1 \mu_2}^{a_1 a_2}(p) 
&=& 
- 
\delta^{a_1 a_2} 
\left( p^2 g_{\mu_1 \mu_2} - p_{\mu_1} p_{\mu_2} \right) J(p^2),
\nn \\
\label{ghost_self-energy}
\widetilde{\Pi}^{a_1 a_2}(p^2) &=& \delta^{a_1 a_2} \frac{ p^2}{ G(p^2)}.
\eea
The ghost-gluon vertex is given by
\bea
\label{ghost-gluon vertex}
\widetilde{\Gamma}_{\mu_3}^{a_1 a_2 a_3}(p_1, p_2; p_3)
\equiv 
-\mbox{i} g \; f^{a_1 a_2 a_3} 
p_1^{\mu}  \widetilde{\Gamma}_{\mu \mu_3}(p_1, p_2; p_3) ,
\eea
where $p_1$ is the out-ghost momentum, $p_2$ is the in-ghost momentum,
$p_3$ and $\mu_3$ are the momentum and the Lorentz index of the gluon
(all momenta are ingoing).

As in Ref.~\cite{DOT1}, we will denote
the zero-loop-order quantities as $X^{(0)}$
and the one-loop-order contributions as $X^{(1)}$,
so that the perturbative expansion is
\[
X=X^{(0)}+X^{(1)}+\ldots \; .
\]
Moreover, at the one-loop level, we  
label the (gauge-dependent) gluon and ghost contributions 
and the (gauge-independent) quark-loop contribution
by the superscripts ``$\xi$'' and ``$q$'', respectively:
\begin{equation}
X^{(1)}=X^{(1,\xi)}+X^{(1,q)}.
\end{equation}
The present paper deals only with quark-loop contributions, $X^{(1,q)}$.
All relevant $X^{(1,\xi)}$ contributions are available 
in \cite{DOT1}.

The lowest-order contributions to $J(p^2)$
and $G(p^2)$ are
\be
J^{(0)}(p^2) = G^{(0)}(p^2) = 1,
\ee
whereas the ghost-gluon vertex at the lowest order is
\bea
\widetilde{\Gamma}^{(0)}_{\mu \mu_3}(p_1,p_2;p_3)=g_{\mu \mu_3}.
\eea
The massive quark-loop contribution, $J^{(1,q)}$,
of the gluon polarization operator is
\be
\label{J^(1,q)}   
J^{(1,q)}(p^2) =
\frac{g^2 \eta}{(4 \pi)^{n/2}}
 \frac{2 \hat{N}_f T_R}{(n - 1) p^2 }
 \biggl\{ [(n - 2) p^2  + 4 m^2] \kappa_{2}(p^2)
- 2 (n - 2) m^2 \widetilde{\kappa} \biggl\} ,
\ee  
where $T_R={\textstyle{1\over8}}\; \mbox{Tr}(I) = {\textstyle{1\over2}}$
(if $\mbox{Tr}(I)=4$), 
with $I$ being the ``unity'' in the space of Dirac matrices.
Furthermore, $\hat{N}_f$ represents a sum over fermions of different 
masses $m$,
\be
\hat{N}_f X(m) \equiv \sum_{i=1}^{N_f} X(m_i).
\ee
In particular, if all fermions have the same mass 
then $\hat{N}_f\Rightarrow N_f$.

The quark-loop contribution to the WST identity 
can be obtained from Eq.~(\ref{ggg_WST}), 
\be
\label{ggg_WST_one-loop_massive}
p_3^{\mu_3} \Gamma^{(1,q)}_{\mu_1 \mu_2 \mu_3}(p_1, p_2, p_3)
=
-\left(g_{\mu_1 \mu_2} p_1^2 - {p_1}_{\mu_1} {p_1}_{\mu_2} \right)
J^{(1,q)}(p_1^2)
+
\left( g_{\mu_1 \mu_2} p_2^2 - {p_2}_{\mu_1} {p_2}_{\mu_2} \right)
J^{(1,q)}(p_2^2),
\ee
where the function $J^{(1,q)}$ is given in Eq.~(\ref{J^(1,q)}),
and we have also taken into account Eq.~(\ref{ghost-gluon vertex}).

\section{Results for the quark-loop contributions}

\subsection{General off-shell case}

The results for the quark-loop contributions to the three-gluon vertex
can be written out in terms of the scalar functions (\ref{J_3 defs}).
It is convenient to define the symmetric combinations:
\begin{eqnarray}
{\cal{Q}} & \equiv &
-\frac{1}{2}(p_1^2+p_2^2+p_3^2)
= (p_1 p_2) + (p_1 p_3) + (p_2 p_3),
\nonumber \\
{\cal{K}} & \equiv & p_1^2 p_2^2 - (p_1 p_2)^2 
= -\frac{1}{4}\lambda(p_1^2,p_2^2,p_3^2),
\label{kkk}
\end{eqnarray}
where $\lambda(x,y,z)=x^2+y^2+z^2-2xy-2yz-2zx$ is the K\"all\'en
function (for other forms of ${\cal K}$,
see Eq.~(3.2) of~\cite{DOT1}).

The general results quoted below have been obtained using the
computer algebra package {\sf REDUCE} \cite{reduce} and standard
techniques\footnote{See in Ref.~\cite{DOT1} for more details.
We note that there is
a misprint in Eq.~(A20) of~\cite{DOT1}, 
$+Z_{010}$ should read $-Z_{010}$.}
for expressing all integrals in terms of a few basic
ones, Eq.~(\ref{J_3 defs}). 

For general values of $p_1^2$, $p_2^2$ and $p_3^2$ we find
the following results
for the scalar functions from Eq.~(\ref{BC-ggg}):
\bea
\label{A-general}
A^{(1,q)}(p_1^2, p_2^2; p_3^2)
&=& \hat{N}_f T_R \frac{g^2 \; \eta}{(4\pi)^{n/2}}
\frac{1}{(n - 1)} 
\bigg[
(n-2) (\kappa_{2,1} + \kappa_{2,2}) 
+4 m^2 \bigg(\frac{\kappa_{2,1}}{p_1^2} 
             + \frac{\kappa_{2,2}}{p_2^2}\bigg)
\nn \\
&&
- 2 (n - 2) m^2 \frac{p_1^2 + p_2^2}{p_1^2  p_2^2} \widetilde{\kappa}
\bigg] ,
\\
\label{B-general}
B^{(1,q)}(p_1^2, p_2^2; p_3^2)
&=& 
\hat{N}_f T_R \frac{g^2 \; \eta}{(4\pi)^{n/2}}
\frac{1}{(n - 1)}
\bigg[
(n-2) (\kappa_{2,1} - \kappa_{2,2}) 
+4 m^2 \bigg( \frac{\kappa_{2,1}}{p_1^2} -
              \frac{\kappa_{2,2}}{p_2^2}\bigg)
\nn \\
&&
+ 2 (n - 2) m^2 \frac{p_1^2 - p_2^2}{p_1^2  p_2^2} \widetilde{\kappa}
\bigg],
\\
\label{C-general}
C^{(1,q)}(p_1^2, p_2^2; p_3^2)
&=& 
\frac{2}{p_1^2-p_2^2} B^{(1,q)}(p_1^2, p_2^2; p_3^2)
,
\\
\label{S-general}
S^{(1,q)}(p_1^2, p_2^2, p_3^2)
&=& 0 ,
\\
\label{F-general}
F^{(1,q)}(p_1^2, p_2^2; p_3^2)
&=&   
-
\hat{N}_f T_R \frac{g^2 \; \eta}{(4\pi)^{n/2}}
\frac{1}{(n-1) (n-2) {\cal{K}}^2}
 \biggl\{
2 (n^2-1) {\cal{K}}^{-1}(p_1 p_2) (p_1 p_3) (p_2 p_3)
\nn \\
&&
\times
 [p_3^2 (p_1 p_2) \varphi_3 \!+\! (p_1 p_3) \kappa_{2,1}
\!+\! (p_2 p_3) \kappa_{2,2} \!+\! p_3^2 \kappa_{2,3}]
\nn \\
&&
+2 (n-1) (n-3) p_3^2 (p_1 p_2)
\big[ (p_1 p_2) \varphi_3+ \kappa_{2,3} \big]
\nn \\
&&
+2 (n-1) (p_3^2-4 m^2)
\left[ 2{\cal K} +3 p_3^2 (p_1 p_2) \right] \varphi_3
\nn \\   
&&
+4 (n-2) (p_3^2-4 m^2)
\left[ (p_1 p_3) \kappa_{2,1} + (p_2 p_3)\kappa_{2,2}
       + p_3^2 \kappa_{2,3} \right]
\nn \\
&&
-2 (3n-5) (p_1 p_2)
\left[ (p_1 p_3) \kappa_{2,1}+ (p_2 p_3) \kappa_{2,2} \right]
\nn \\
&&
+\left[ n (n-4) {\cal K} - 2(n+1)p_3^2 (p_1 p_2) \right]
( \kappa_{2,1}+\kappa_{2,2} )
\nn \\
&&
+ 2 (n-2) \frac{m^2{\cal K}}{p_1^2 p_2^2}
\left[  (p_3^2 - 2 p_1^2) \kappa_{2,1}
      + (p_3^2 - 2 p_2^2) \kappa_{2,2}
- 2 (n-2) (p_1 p_2) \widetilde{\kappa} \right]
\nn \\
&&
+(n-2){\cal K} (p_1-p_2)^2
\biggl( n-2 + 2 m^2 \frac{p_1^2+ p_2^2}{p_1^2 p_2^2} \biggl)
\frac{\kappa_{2,1}-\kappa_{2,2}}{p_1^2-p_2^2}
\biggl\} \; ,
\\
\label{H-general}
H^{(1,q)}(p_1^2, p_2^2, p_3^2)
&=&
-
\hat{N}_f T_R \frac{g^2 \; \eta}{(4\pi)^{n/2}}
\frac{2}{(n-1) (n-2) {\cal{K}}^2}
 \bigg\{
(n^2-1) {\cal{K}}^{-1}(p_1   p_2) (p_1   p_3) (p_2   p_3)
\nn \\
&&
\times
[(p_1   p_2) (p_1   p_3)
 (p_2   p_3) \varphi_3
+ (p_1   p_2) (p_1   p_3) \kappa_{2,1}
\nn \\
&&
\hspace*{4.2cm} + \, (p_1 p_2)(p_2   p_3) \kappa_{2,2}
 + (p_1   p_3) (p_2   p_3) \kappa_{2,3}]
\nn \\
&&
- 3 (n-1) (p_1   p_2) (p_1   p_3) (p_2   p_3) \left[
( {\cal{Q}} + 4m^2 )
\varphi_3 + \kappa_{2,1} + \kappa_{2,2} + \kappa_{2,3} \right]
\nn \\
&&
+(n-1) (n-2) {\cal{K}}^2 \varphi_3
- 2 (n-2)^2 {\cal{K}} m^2 \widetilde{\kappa}
\nn \\
&&
+(n-2) (p_1^2-4m^2) 
[p_1^2 (p_2 p_3) + (p_1 p_2) (p_1 p_3)] \kappa_{2,1}
\nn \\
&&
+(n-2) (p_2^2-4m^2)
[p_2^2 (p_1 p_3) + (p_1 p_2) (p_2 p_3)] \kappa_{2,2}
\nn \\
&&
+(n-2) (p_3^2-4m^2)
[p_3^2 (p_1 p_2) + (p_1 p_3) (p_2 p_3)] \kappa_{2,3}
\bigg\} \; .
\eea
We note that $A^{(1,q)}$, $B^{(1,q)}$ and $C^{(1,q)}$
do not depend on the third argument, $p_3^2$.
The function $H^{(1,q)}(p_1^2,p_2^2,p_3^2)$ is explicitly
symmetric with respect to all arguments.
Once more, we would like to note that the $S$ function vanishes
at the one-loop level.

In the limit $n\to 4$ ($\ep\to 0$), only the $A^{(1,q)}$
function has a $1/\ep$ singularity of an ultraviolet origin,
\be
\frac{g^2\eta}{(4\pi)^{2-\ep}}\;
\frac{4}{3} N_f T_R \left( \frac{1}{\ep} + \ldots \right) .
\ee
It can be renormalized by the corresponding 
renormalization factor, $Z_1$.

To consider the massless limit $m=0$ of these functions, 
we should put ${\widetilde{\kappa}} \Rightarrow 0$
(massless tadpole), 
$\varphi_3 \Rightarrow \varphi_0 \equiv \varphi$,
$\kappa_{2,i} \Rightarrow \kappa_{0,i} \equiv \kappa_i$.
In this way, we reproduce Eqs.~(3.17)--(3.22) of
Ref.~\cite{DOT1}. 

\subsection{Symmetric case, \boldmath $p_1^2=p_2^2=p_3^2\equiv p^2$}

In the symmetric limit $p_1^2=p_2^2=p_3^2 \equiv p^2$
[$(p_1 p_2)=(p_1 p_3)=(p_2 p_3)=-\frac{1}{2} p^2$],  
the three-gluon vertex 
has only three independent tensor structures \cite{CelGon},
\bea
\label{sym_vertex}
\Gamma_{\mu_1\mu_2\mu_3}(p_1,p_2,p_3)
&=&            \left[ g_{\mu_1 \mu_2}(p_1-p_2)_{\mu_3}
                  + g_{\mu_2 \mu_3}(p_2-p_3)_{\mu_1}
                  + g_{\mu_3 \mu_1}(p_3-p_1)_{\mu_2} \right] G_0(p^2)
\nn \\
&&  
- (p_2-p_3)_{\mu_1} (p_3-p_1)_{\mu_2} (p_1-p_2)_{\mu_3} G_1(p^2)
\nn \\
&&
+
\left( {p_2}_{\mu_1} {p_3}_{\mu_2} {p_1}_{\mu_3}
-  {p_3}_{\mu_1} {p_1}_{\mu_2} {p_2}_{\mu_3}  \right) G_2(p^2).
\eea
The $G_i$ functions are related to the scalar functions in
Eq.~(\ref{BC-ggg}) via (see in~\cite{DOT1})
\bea
G_1(p^2)
&=& C(p^2,p^2;p^2) + \frac{1}{2}  p^2 F(p^2,p^2;p^2),
\nn \\
G_2(p^2)
&=& G_1(p^2) + H(p^2,p^2,p^2),
\nn \\
G_0(p^2)
&=& A(p^2,p^2;p^2) + \frac{1}{2} p^2 G_2(p^2) .
\eea
Note that in the symmetric case the $B$ and $S$ functions 
vanish at any order, because they are antisymmetric.

The one-loop quark contributions to the $G_i$ functions 
can be obtained directly from the general expressions
(\ref{A-general})--(\ref{H-general}), using
the substitution (see Appendix~C in~\cite{DOS})
\bea
\label{pp1pp2}
&& \hspace*{-10mm}
\frac{\kappa_{2}(p_2^2)-\kappa_{2}(p_1^2)}
{p_2^2-p_1^2}\bigg|_{p_1^2=p_2^2=p^2}
= \frac{\mbox{d}\kappa_{2}(p^2)}{\mbox{d}p^2}
\nn \\
&&
= \frac{1}{2 p^2(p^2-4m^2)}
\left\{
[(n-4)p^2+4m^2] \kappa_{2}(p^2)
-2(n-2) m^2\widetilde\kappa \right\} .
\eea
In this way, we obtain
\bea
\label{g_i_functions_q}
G_0^{(1,q)}(p^2)
&=& -\hat{N}_f T_R \frac{g^2 \; \eta}{(4\pi)^{n/2}}
\frac{2}{9 (n - 2)}
 \bigg\{ 2 [(3 n - 8) p^2+6 m^2] \varphi_{3 {\rm s}}
- 3 (3 n - 8) \kappa_{2}(p^2)\bigg\} ,
\nn \\
G_1^{(1,q)}(p^2)
&=& -\hat{N}_f T_R \frac{g^2 \; \eta}{(4\pi)^{n/2}}
 \frac{4}{27 p^2}
 \bigg\{ 4 p^2 
\varphi_{3 {\rm s}}
+ 3 \frac{n \!-\! 4}{n \!-\! 1}
\kappa_{2}(p^2)
+\frac{18 m^2}{(n \!-\! 1) p^2}
 [2 \kappa_{2}(p^2) \!-\! (n \!-\!2) \widetilde{\kappa}]\bigg\} ,
\nn \\
G_2^{(1,q)}(p^2)
&=& -\hat{N}_f T_R \frac{g^2 \; \eta}{(4\pi)^{n/2}}
\frac{4}{9 p^2}
 \bigg\{
\frac{2}{n - 2}
 [(3 n - 8) p^2 + 6 m^2] 
\varphi_{3 {\rm s}}
\nn \\
&&
- 3 \frac{n - 4}{(n - 1) (n - 2)} \kappa_{2}(p^2)
+ \frac{18 m^2}{(n-1) p^2} [2 \kappa_{2}(p^2) 
- (n - 2) \widetilde{\kappa}]\bigg\} ,
\eea
where $\varphi_{3 {\rm s}}\equiv \varphi_3(p^2,p^2,p^2)$
is the three-point function in the symmetric limit
whose analytic properties are discussed in detail
in Sec.~4.2.
In the massless quark limit we reproduce the corresponding results 
(3.33)--(3.35) of~\cite{DOT1}.

We can also consider the limit $p^2\to0$, when all
external gluons are on shell.
Since there are some $p^2$
in the denominators of the r.h.s.'s of Eqs.~(\ref{g_i_functions_q}),
we should keep a few terms of the expansion\footnote{Any number  
of terms of the small-$p^2$ expansion of $\kappa_2(p^2)$ and
$\varphi_3(p^2,p^2,p^2)$
can be obtained from Eqs.~(17), (37) and (38) of \cite{BD-TMF}.}
in $p^2$,
\begin{equation}
\label{kappa2_exp}
\kappa_2(p^2) = \frac{1}{2}(n-2){\widetilde{\kappa}}
\left[ 1 - \frac{n-4}{12}\; \frac{p^2}{m^2}
+\frac{(n-4)(n-6)}{240} \; \frac{(p^2)^2}{m^4} 
+ {\cal O}\left( \frac{(p^2)^3}{m^6} \right)
\right] ,
\end{equation}
\begin{equation}
\label{phi3_exp1}
\varphi_{3 {\rm s}} =
\varphi_3(p^2,p^2,p^2) =
\frac{(n-4)(n-2)}{8 m^2} {\widetilde{\kappa}}
\left[ 1 - \frac{n-6}{8}\; \frac{p^2}{m^2}
+ {\cal O}\left( \frac{(p^2)^2}{m^4} \right)
\right] .
\end{equation}
In this way, we obtain
\bea
\label{G_i(0)}
G_0^{(1,q)}(0) &=& 
\hat{N}_f T_R \frac{g^2\eta}{(4\pi)^{n/2}}\;
\frac{2}{3} (n-2) {\widetilde{\kappa}} ,
\nn \\
G_1^{(1,q)}(0) &=& - \hat{N}_f T_R \frac{g^2\eta}{(4\pi)^{n/2}}\;
\frac{1}{15 m^2} (n-4) (n-2) {\widetilde{\kappa}} ,
\nn \\
G_2^{(1,q)}(0) &=& - \hat{N}_f T_R \frac{g^2\eta}{(4\pi)^{n/2}}\;  
\frac{17}{60 m^2} (n-4) (n-2) {\widetilde{\kappa}} .
\eea
Therefore, the three-gluon vertex is in the totally on-shell
case given
by Eq.~(\ref{sym_vertex}) with the functions (\ref{G_i(0)}).

\subsection{One gluon on shell, \boldmath $p_3^2=0$}
\label{One gluon on shell}

In the limit $p_3^2 = 0$ 
[$(p_1 p_2) = -\frac{1}{2}(p_1^2+p_2^2)$], 
the tensor structures in Eq.~(\ref{BC-ggg}) remain unchanged.  
The results for 
the scalar functions of arguments $(p_1^2, p_2^2, 0)$ can be
obtained from the general expressions 
(\ref{A-general})--(\ref{H-general}).
In fact, the results (\ref{A-general})--(\ref{C-general})
do not depend on $p_3^2$. Therefore, their form is not
changed. In the expressions (\ref{F-general}) 
and (\ref{H-general}), we should
put $\kappa_{2,3}\Rightarrow \frac{1}{2}(n-2){\widetilde{\kappa}}
(1 - \frac{n-4}{12}\; \frac{p_3^2}{m^2})$
[cf.\ Eq.~(\ref{kappa2_exp})] and
\be
\label{one_os_phi}
\varphi_3(p_1^2,p_2^2,0) 
= \frac{p_1^2 \varphi_3(p_1^2,0,0) - p_2^2 \varphi_3(0,p_2^2,0)}
       {p_1^2-p_2^2}
\ee
[see Eq.~(\ref{one_os_}) below]. In this way, we get
\bea
\label{ABCFH_pp1_pp2_0}
F^{(1,q)}(p_1^2, p_2^2; 0)
&=& 
\hat{N}_f T_R \frac{g^2 \; \eta}{(4\pi)^{n/2}} 
\frac{4}{(n - 1) (n-2) (p_1^2 - p_2^2)^2}
 \bigg[ 
n (n-2) (\kappa_{2,1}+\kappa_{2,2}) 
\nn \\
&&
-16 (n - 1) m^2\frac{p_1^2 \varphi_3(p_1^2,0,0)-p_2^2 
\varphi_3(0,p_2^2,0) }{p_1^2 - p_2^2}
-4 (n-2) m^2
\bigg(\frac{\kappa_{2,1}}{p_1^2}+\frac{\kappa_{2,2}}{p_2^2}\bigg) 
\nn \\
&&
- 4 n \frac{p_1^2 \kappa_{2,1}-p_2^2 \kappa_{2,2}}{p_1^2 - p_2^2}
-16 (n-2) m^2 \frac{\kappa_{2,1}-\kappa_{2,2}}{p_1^2 - p_2^2}
+ 2 (n - 2)^2 m^2 \frac{p_1^2 + p_2^2}{p_1^2 p_2^2} \widetilde{\kappa}
\bigg], 
\nn \\
H^{(1,q)}(p_1^2, p_2^2, 0)
&=& 
 \hat{N}_f T_R \frac{g^2 \; \eta}{(4\pi)^{n/2}} 
\frac{2}{(n - 1) (n - 2)
(p_1^2 - p_2^2)^3} 
\nonumber \\
&&
\times \biggl\{
4 (n - 1) [(n - 2) p_1^2 p_2^2 + 6 m^2 (p_1^2 + p_2^2)] 
[p_1^2 \varphi_3(p_1^2,0,0)-p_2^2 \varphi_3(0,p_2^2,0)] 
\nonumber \\
&&
+2(n^2 - 1) (p_1^2  + p_2^2 )^2(\kappa_{2,1} -\kappa_{2,2})
+ 6 (n - 1) [(p_1^2) ^2 -(p_2^2)^2](\kappa_{2,1} + \kappa_{2,2})
\nonumber \\
&&
+ 4 (n - 2) [(4 m^2 - p_1^2 ) (3 p_1^2  + p_2^2 ) \kappa_{2,1} 
           - (4 m^2 - p_2^2 ) (p_1^2  + 3 p_2^2 ) \kappa_{2,2}]
\nonumber \\ 
&&
- (n - 2)^2 [(n - 1) (p_1^2 + p_2^2) + 16 m^2] (p_1^2 - p_2^2) 
\widetilde{\kappa} 
\biggl\}.
\eea
For the functions with permuted arguments 
$(0, p_1^2,p_2^2)$, we obtain
\bea
\label{ABCF_0_pp1_pp2}
A^{(1,q)}(0, p_1^2; p_2^2)
&=& 
\hat{N}_f T_R \frac{g^2 \; \eta}{(4\pi)^{n/2}} 
 \frac{1}{n - 1} 
\bigg\{ 
\frac{1}{3} (n - 2) 
[3 \kappa_{2,1} + (n - 1) \widetilde{\kappa}] 
+ 2 \frac{m^2}{p_1^2} [2 \kappa_{2,1} - (n - 2) \widetilde{\kappa}]
\bigg\}
,
\nn \\ 
B^{(1,q)}(0, p_1^2; p_2^2)
&=& 
- \hat{N}_f T_R \frac{g^2 \; \eta}{(4\pi)^{n/2}} 
 \frac{1}{n - 1} 
\bigg\{ 
\frac{1}{3} (n - 2) 
[3 \kappa_{2,1} - (n - 1) \widetilde{\kappa}] 
+ 2 \frac{m^2}{p_1^2} [2 \kappa_{2,1} - (n - 2) \widetilde{\kappa}]
\bigg\}
,
\nn \\
C^{(1,q)}(0, p_1^2; p_2^2)
&=& - \frac{2}{p_1^2} B^{(1,q)}(0, p_1^2; p_2^2)
,
\nn \\
F^{(1,q)}(0, p_1^2; p_2^2)
&=&
-\hat{N}_f T_R \frac{g^2 \; \eta}{(4\pi)^{n/2}} \frac{4}{(n - 1) (n - 2)
(p_1^2 - p_2^2)^3} 
\nn \\ 
&&
\times
\bigg\{
4 (n-1) 
[(n-2) p_1^2 p_2^2+4 m^2 (p_1^2+2 p_2^2)] \frac{p_1^2 
\varphi_3(p_1^2,0,0)-p_2^2 \varphi_3(0,p_2^2,0)}{p_1^2-p_2^2} 
\nn \\ 
&&
+8 p_2^2 
[
(n-1)^2 p_1^2
+2 p_2^2
+8 (n-2) m^2 
]
\frac{\kappa_{2,1}-\kappa_{2,2}}{p_1^2 - p_2^2}
\nn \\ 
&&
 + 2 [(n+2) p_1^2-(n-10) p_2^2] \kappa_{2,1}
-(n - 1) (n - 2)^2 p_2^2 [3- p_2^2 (p_1^2)^{-1} ] \widetilde{\kappa} 
\nn \\
&&
- \frac{1}{3}(n-2)^2 (p_1^2)^{-1} (p_1^2-p_2^2)^2 [3 \kappa_{2,1}
+2(n-1) \widetilde{\kappa}]
\nn \\ 
&&
+ 2 (n-2) m^2(p_1^2)^{-2} (5 p_1^2 - p_2^2) (p_1^2+p_2^2) 
[2 \kappa_{2,1}-(n-2)\widetilde{\kappa}]
\bigg\}.
\eea
The functions $A$, $B$, $C$ and $F$ with the arguments $(p_2^2,0, p_1^2)$
can be obtained from those with arguments $(0, p_1^2, p_2^2)$
by interchanging $p_1^2 \leftrightarrow p_2^2$. We remind that 
$A$, $C$ and $F$ are symmetric in the first two
arguments, whereas $B$ is antisymmetric. The function $H$ is totally
symmetric with respect to all the three arguments.

Considering the massless limit, we should substitute
${\widetilde{\kappa}}\Rightarrow~0$, and
\begin{equation}
\varphi_3(p_1^2,0,0)\Rightarrow
-\frac{2(n-3)\kappa_1}{(n-4)p_1^2}, \qquad
\varphi_3(0,p_2^2,0)\Rightarrow
-\frac{2(n-3)\kappa_2}{(n-4)p_2^2}.
\end{equation}
Using these relations, we see that the results given in 
Eqs.~(\ref{ABCFH_pp1_pp2_0}) and (\ref{ABCF_0_pp1_pp2})
[together with Eqs.~(\ref{A-general})--(\ref{C-general})]
agree with Eqs.~(4.15)--(4.23) given in Ref.~\cite{DOT1}.
We have also checked that in the massive case
Eq.~(4.24) of~\cite{DOT1} is satisfied by the one-loop expressions.

\subsection{Zero-momentum limit, \boldmath $p_3=0$}

Putting $p_3=0$ ($p_1=-p_2\equiv p$, $p_1^2=p_2^2=p^2$), 
we get the three-gluon vertex Eq.~(\ref{BC-ggg})
in terms of three tensor structures \cite{BF,DOT1},
\bea
\label{Gamma_p30}
\Gamma_{\mu_1 \mu_2 \mu_3}(p,-p,0)
&=& 2 g_{\mu_1 \mu_2} p_{\mu_3}  
\left[ A(p^2, p^2; 0) + p^2  C(p^2, p^2; 0) \right]
- 2 p_{\mu_1} p_{\mu_2} p_{\mu_3} C(p^2, p^2; 0)   
\nn \\
&&
- \left( g_{\mu_1 \mu_3} p_{\mu_2} + g_{\mu_2 \mu_3} p_{\mu_1} \right)
  \left[ A(0, p^2; p^2) - B(0, p^2; p^2) \right] .
\eea
Moreover, due to the relation 
\be
\label{aba_rel_p30}
  A(0, p^2; p^2)
- B(0, p^2; p^2)
- A(p^2, p^2; 0)
= 0 ,  
\ee   
the tensor structures in Eq.~(\ref{Gamma_p30}) 
reduce to two independent ones \cite{BL,DOT1},
\bea
\label{Gamma_p30_2_ind}
\Gamma_{\mu_1 \mu_2 \mu_3}(p, -p, 0)
&=& \left( 2 g_{\mu_1 \mu_2} p_{\mu_3} - g_{\mu_1 \mu_3} p_{\mu_2}
         - g_{\mu_2 \mu_3} p_{\mu_1} \right) A(p^2, p^2; 0) 
\nn \\
&&
 + 2 p_{\mu_3} \left( p^2 g_{\mu_1 \mu_2} - p_{\mu_1} p_{\mu_2} \right)
                                               C(p^2, p^2; 0) .
\eea   
This reduction is a corollary of the differential (zero-momentum)
version of the WST identity. It is discussed in detail in Sec.~III
of Ref.~\cite{DOT2}. Note that 
the functions $T_i(p^2)$ used in Refs.~\cite{BL,DOT2}
are related to the functions in Eq.~(\ref{Gamma_p30_2_ind}) as
\be
T_1(p^2) \leftrightarrow A(p^2,p^2;0), \quad
T_2(p^2) \leftrightarrow -2p^2 C(p^2,p^2;0) \; .
\ee

The results for these functions can be obtained from
Eqs.~(\ref{A-general}) and (\ref{C-general})
by taking the limit 
$p_2^2\to p_1^2\equiv p^2$, using Eq.~(\ref{pp1pp2}). 
In this way, we obtain
\bea
\label{A_ppC_p30}
A^{(1,q)}(p^2, p^2; 0)
&=&
\hat{N}_f T_R \frac{g^2 \; \eta}{(4\pi)^{n/2}} \frac{2}{n - 1}
 \bigg[\bigg(n - 2+ 4 \frac{m^2}{p^2}\bigg) \kappa_2(p^2)
- 2 (n - 2) \frac{m^2}{p^2} \widetilde{\kappa}\bigg] ,
\nn \\
C^{(1,q)}(p^2, p^2; 0)
&=& 
\hat{N}_f T_R \frac{g^2 \; \eta}{(4\pi)^{n/2}} \frac{1}{(n - 1) (p^2-4 m^2)}
 \bigg[ (n - 4) \bigg(n - 2+ 4 \frac{m^2}{p^2}\bigg) \kappa_2(p^2) 
\nn \\
&&
+2 \frac{m^2}{p^2} \bigg(n- 4 + 12 \frac{m^2}{p^2}\bigg)
 [ 2\kappa_2(p^2)-(n - 2) \widetilde{\kappa}]
\bigg] \; .
\eea
In the massless limit, putting 
$\kappa_2(p^2)\Rightarrow\kappa_0(p^2)\equiv\kappa(p^2)$, 
${\widetilde{\kappa}}\Rightarrow0$, we see that 
the results given in Eq.~(\ref{A_ppC_p30})
agree with Eqs.~(4.32)--(4.34) of Ref.~\cite{DOT1}. 

\subsection{Two gluons on shell, \boldmath $p_1^2=p_2^2=0$}

In the limit $p_1^2=p_2^2=0$ 
[$p_3^2 \equiv p^2$, $(p_1 p_2)=\frac{1}{2} p^2$],
the three-gluon vertex (\ref{BC-ggg}) involves
seven tensor structures, and thus seven independent 
scalar functions $U_i(p^2)$ (see in Refs.~\cite{BF,DOT1}),
\begin{eqnarray}
\label{ggg_pp12=0}
\left. 
\Gamma_{\mu_1 \mu_2 \mu_3}(p_1, p_2, p_3)\right|_{p_1^2=p_2^2=0}
&=& g_{\mu_1 \mu_2} (p_1-p_2)_{\mu_3}  U_1(p^2)
+ \left( g_{\mu_1 \mu_3} {p_1}_{\mu_2} - g_{\mu_2 \mu_3} {p_2}_{\mu_1} 
  \right) U_2(p^2)
\nonumber \\
&&
+\left( g_{\mu_1 \mu_3} {p_2}_{\mu_2} - g_{\mu_2 \mu_3} {p_1}_{\mu_1}
  \right) U_3(p^2)
+ {p_1}_{\mu_1} {p_2}_{\mu_2} (p_1-p_2)_{\mu_3}  U_4(p^2)
\nonumber \\
&&
+ {p_1}_{\mu_2} {p_2}_{\mu_1} (p_1-p_2)_{\mu_3} U_5(p^2)
\nonumber \\
&&
+ \left({p_1}_{\mu_1} {p_1}_{\mu_2} {p_1}_{\mu_3} 
        -{p_2}_{\mu_1} {p_2}_{\mu_2} {p_2}_{\mu_3} \right)  U_6(p^2)
\nonumber \\
&&
+ \left( {p_1}_{\mu_1} {p_1}_{\mu_2} {p_2}_{\mu_3}
     -{p_2}_{\mu_1} {p_2}_{\mu_2} {p_1}_{\mu_3} \right) U_7(p^2) .
\end{eqnarray}
The functions $U_i(p^2)$
can be related to the functions $A$, $B$, $C$, $F$ and $H$ 
of Eq.~(\ref{BC-ggg}). The corresponding relations are
given in Eqs.~(4.59)--(4.65) of Ref.~\cite{DOT1}.

The results for one-loop contributions to the 
scalar functions can be obtained in two ways.
First, taking the limit $p_1^2=p_2^2 \equiv p^2$ 
and then putting $p^2=0$.
Secondly, starting from the results of 
Sec.~\ref{One gluon on shell} and thereafter putting
another momentum on shell. Both ways lead to the same  
results for quark-loop contributions to the $U_i$ functions:
\bea
\label{U_i}
U_1^{(1,q)}(p^2) &=& 
-\hat{N}_f T_R \frac{g^2 \; \eta}{(4\pi)^{n/2}}\frac{2}{(n - 1) (n - 2)}
 \bigg\{
4 (n - 1) m^2  \varphi_3(0,0,p^2) - n (n - 3) \kappa_2(p^2)
\nonumber \\
&&
+4 (n - 2) \frac{m^2}{p^2} [2 \kappa_2(p^2)-(n - 2) 
\widetilde{\kappa}]
\bigg\} ,
\nonumber \\ 
U_2^{(1,q)}(p^2) &=& 
-\hat{N}_f T_R \frac{g^2 \; \eta}{(4\pi)^{n/2}}\frac{4}{(n - 1)}
 \bigg\{
(n - 2) \kappa_2(p^2) 
+2 \frac{m^2}{p^2} 
[2 \kappa_2(p^2)-(n - 2) \widetilde{\kappa}]
\bigg\} ,
\nonumber \\ 
U_3^{(1,q)}(p^2) &=& 
\hat{N}_f T_R \frac{g^2 \; \eta}{(4\pi)^{n/2}}
\frac{1}{(n - 1) (n - 2)}
\bigg\{
 8 (n - 1) m^2 \varphi_3(0,0,p^2) +4 \kappa_2(p^2) 
\nonumber \\
&&
+8 (n - 2) \frac{m^2}{p^2} [2 \kappa_2(p^2)-(n - 2) 
\widetilde{\kappa}]
- (n -1) (n - 2)^2 \widetilde{\kappa}
\bigg\} ,
\nonumber \\ 
U_4^{(1,q)}(p^2) &=& 
\hat{N}_f T_R \frac{g^2 \; \eta}{(4\pi)^{n/2}}
 \frac{2}{(n - 1) (n - 2) p^2}
 \bigg\{
 16 (n - 1) m^2 \varphi_3(0,0,p^2) + 2 (n + 2) \kappa_2(p^2)
\nonumber \\
&&
+12 (n - 2) \frac{m^2}{p^2} [2 \kappa_2(p^2)-(n - 2)
\widetilde{\kappa}] 
-(n - 1) (n - 2)^2 \widetilde{\kappa}
\bigg\} ,
\nonumber \\ 
U_5^{(1,q)}(p^2) &=& 
\hat{N}_f T_R \frac{g^2 \; \eta}{(4\pi)^{n/2}}\frac{4}{(n - 1) (n - 2) p^2}
 \bigg\{
4 (n - 1) m^2 \varphi_3(0,0,p^2) - (n - 4) \kappa_2(p^2)
\nonumber \\
&&
+6 (n - 2) \frac{m^2}{p^2} [2 \kappa_2(p^2)-(n - 2) 
\widetilde{\kappa}]
\bigg\} ,
\nonumber \\
U_6^{(1,q)}(p^2)
&=& 
\hat{N}_f T_R \frac{g^2 \; \eta}{(4\pi)^{n/2}} 
\frac{2}{(n-1) p^2}
\Biggl\{ \left( n-2+\frac{4m^2}{p^2} \right)
[2 \kappa_2(p^2) -(n -2) \widetilde{\kappa}]
\nonumber \\
&&
+ \frac{1}{3} (n-2)(n-4) \widetilde{\kappa}
\Biggl\} ,
\nonumber \\ 
U_7^{(1,q)}(p^2) &=& 
\hat{N}_f T_R \frac{g^2 \; \eta}{(4\pi)^{n/2}}
\frac{2}{(n - 1) (n - 2) p^2}
 \bigg\{
8 (n - 1) m^2 \varphi_3(0,0,p^2) - 2 (n - 4) \kappa_2(p^2)
\nonumber \\
&&
+ (n - 2) \bigg(n - 1 + 12 \frac{m^2}{p^2}\bigg) 
[2\kappa_2(p^2)-(n - 2) \widetilde{\kappa}]
\bigg\} .
\eea
In the massless limit $m=0$, we should put 
$\widetilde{\kappa}\Rightarrow0$. In this way,
we reproduce the corresponding results (F8)--(F14) of Ref.~\cite{DOT1}.

A useful check on the results (\ref{U_i}) is to consider
the limit $p^2\to0$, when the third gluon is also on shell.
Since we have some $p^2$ in the denominators, we need to
take a few terms of the expansions of $\kappa_2(p^2)$ and
$\varphi_3(0,0,p^2)$ in $p^2$. The expansion of $\kappa_2(p^2)$
is given in Eq.~(\ref{kappa2_exp}), while (cf.\ Eqs.~(37), (38)
of~\cite{BD-TMF})
\be
\label{phi3_exp2}
\varphi_3(0,0,p^2) =  
\frac{(n-4)(n-2)}{8 m^2} {\widetilde{\kappa}}
\left[ 1 - \frac{n-6}{24}\; \frac{p^2}{m^2}
+ {\cal O}\left( \frac{(p^2)^2}{m^4} \right)
 \right] .
\ee
In this way, we reproduce the results (\ref{sym_vertex}), 
(\ref{G_i(0)}) for the totally on-shell configuration.

\section{Scalar three-point function}

\subsection{General off-shell case}

We collect here some results for
the scalar three-point integrals~(\ref{defJ3}).
General results for such integrals are given 
in~\cite{BD-TMF}\footnote{There is a misprint in  
a representation for the $\Phi_3$ function given in the
last line of Eq.~(38) of~\cite{BD-TMF}: the arguments
of the hypergeometric function should read
$\frac{z_1}{4}, \frac{z_2}{4}, \frac{z_3}{4}$
(rather than $z_1, z_2, z_3$). 
The representation given in 
the second line of Eq.~(38) of~\cite{BD-TMF}, as well as 
the generalization of this result to the case of $N$-point 
diagrams given in Eq.~(4.7) of~\cite{jmp1}, are correct.},
in terms of a triple hypergeometric series
in the variables $p_i^2/m^2$.

We are mainly interested in the case of unit powers 
of the denominators, $\nu_1=\nu_2=\nu_3=1$.
We note that all integrals with higher integer
powers of the propagators can be reduced to $J_3(1,1,1)$
and two-point integrals, by using a recurrence 
procedure~\cite{JPA} based on the integration-by-parts
technique~\cite{ibp}.

Transforming Feynman parametric integrals, we get
\be
\label{2fold3}
J_3(1,1,1) 
=
-{\mbox{i}}\pi^{n/2} \Gamma\biggl(3-\frac{n}{2}\biggl)
\int\limits_0^{\infty} \int\limits_0^{\infty}
\frac{{\mbox{d}}\xi\;\; {\mbox{d}}\eta}
{(1\!+\!\xi\!+\!\eta)^{n-3}
\left[m^2 (1\!+\!\xi\!+\!\eta)^2 \!-\!\xi  p_1^2
      \!-\!\eta p_2^2\!-\!\xi\eta p_3^2\right]^{3-n/2}} \; .
\ee
In three dimensions ($n=3$), these integrals can be evaluated
in terms of elementary functions~\cite{Nickel} (see also
Sec.~VA in~\cite{DD}). 
In four dimensions, shifting the integration variables 
in~(\ref{2fold3}), one can obtain
the standard representation of the three-point function
in terms of dilogarithms~\cite{tHV'79} (see also 
in~\cite{vOV}). 

Another representation of the four-dimensional result,
in terms of the Clausen function 
$\Cl{2}{\theta}={\rm Im}\left[\Li{2}{e^{{\rm i}\theta}}\right]$
(see in~\cite{Lewin}), can be derived using
the geometrical approach of~\cite{DD} (see also in~\cite{Wagner}).
Further details and explicit results for the general case
can be found in Sec.~VB of~\cite{DD}. 
These results are related to those of Ref.~\cite{tHV'79}
by analytic continuation.

We would like to note that the approach of~\cite{DD} 
also provides results valid in an arbitrary dimension
$n=4-2\ep$, 
in terms of one-fold angular integrals 
of the type\footnote{For the general three-point function with
different masses, we get six integrals 
of the type~(\ref{DDangular}). In the case of equal masses,
the number of different integrals reduces to three 
(cf.\ Fig.~7 of Ref.~\cite{DD}).}
\be
\label{DDangular}
\frac{1}{2\ep} \int\limits_0^{\varphi/2}
{\rm d}\phi \left[ 1-
\left(1+\frac{\tan^2\eta}{\cos^2\phi}\right)^{-\ep}
\right]
\ee
(see also in~\cite{Crete,DK}). Using a simple
substitution of variables, $\phi=\arctan(\sqrt{z}/\sin\eta)$, 
the results for these integrals can be expressed
in terms of Appell's hypergeometric function $F_1$ of 
two variables, similar to those obtained
by Tarasov~\cite{bastei_tar} by using recurrence relations
with respect to the space-time dimension. 

  The structure of singularities of the general three-point
  function, including the anomalous thresholds, was studied
  in Ref.~\cite{singularities}.

\subsection{Symmetric case}

In the completely symmetric case, we have 
$p_1^2=p_2^2=p_3^2\equiv p^2$, $m_1=m_2=m_3\equiv m$. 
Let us follow the geometrical approach of~\cite{DD}
to calculate the integral $J_3(1,1,1)$ in this case.
The geometrical variables defined in Fig.~6 of~\cite{DD}
in this symmetric configuration take
the following values:
\bea
&& \tau_{12}=\tau_{23}=\tau_{31} \equiv \tau, \quad
\cos\tau = 1-\frac{p^2}{2m^2},
\\
&& D^{(3)}= (1-\cos\tau)^2 (1+2\cos\tau)
= \frac{(p^2)^2}{4m^4}\left(3-\frac{p^2}{m^2}\right),
\\
&& \Lambda^{(3)} = \frac{3}{4}(p^2)^2, \quad
m_0 = m^3 \sqrt{\frac{D^{(3)}}{\Lambda^{(3)}}}
= m \sqrt{1-\frac{p^2}{3m^2}},
\\
&& \tau_{01} = \tau_{02} = \tau_{03} \equiv \tau_0, \quad
\cos\tau_0 = \frac{m_0}{m} = \sqrt{1-\frac{p^2}{3m^2}},
\\
&& \varphi_{12} = \varphi_{23} = \varphi_{31} \equiv \varphi
= \frac{2\pi}{3},
\\
&& \eta_{12} = \eta_{23} = \eta_{31} \equiv \eta, \quad
\tan\eta = \cos\frac{\varphi}{2} \tan\tau_0
= \frac{1}{2}\tan\tau_0,
\\
&& \tan^2\eta = \frac{p^2}{4(3m^2-p^2)} \; .
\eea
The latter quantity is positive for $0<p^2<3m^2$,
and negative otherwise.

Let us use Eqs.~(5.16)--(5.17) of~\cite{DD}, remembering
that in the symmetric case the result should be
multiplied by
\be
-\frac{3 {\rm i} \pi^2}{\sqrt{\Lambda^{(3)}}}
= - \frac{2 {\rm i} \pi^2 \sqrt{3}}{p^2} \; .
\ee
{}From Eq.~(5.16) of~\cite{DD} [cf.\ also Eq.~(\ref{DDangular})]
we get, for the case of four dimensions:
\be
\label{DD(5.16)sym}
J_3(1,1,1)\bigr|_{p_i^2=p^2, \; n=4}
= -\frac{2 {\rm i} \pi^2 \sqrt{3}}{p^2}
\int\limits_0^{\pi/3} {\rm d}\phi \;
\ln\left(1+\frac{\tan^2\eta}{\cos^2\phi}\right) \; .
\ee

In the region $0<p^2<3m^2$ (when the momentum is timelike
but its square does not exceed the anomalous threshold $3m^2$),
we can directly use the geometrical result (5.17) 
from~\cite{DD},
\bea 
\label{DD(5.17)sym}
J_3(1,1,1)\bigr|_{p_i^2=p^2, \; n=4}
&=& -\frac{{\rm i} \pi^2 \sqrt{3}}{p^2}
\Biggl[ \Cl{2}{\frac{2\pi}{3}+\tau}
+\Cl{2}{\frac{2\pi}{3}-\tau}-2\Cl{2}{\frac{2\pi}{3}}
\nonumber \\ &&
+ \tau \ln\left(
\frac{\sin\left(\frac{\pi}{3}+\frac{\tau}{2}\right)}
     {\sin\left(\frac{\pi}{3}-\frac{\tau}{2}\right)} \right)
\Biggl] ,
\eea
where $\Cl{2}{\theta}$ is the Clausen function \cite{Lewin}
[remember that 
$\Cl{2}{\frac{2\pi}{3}}=\frac{2}{3}\Cl{2}{\frac{\pi}{3}}$].
Note that there is a logarithmic singularity at
the anomalous threshold $p^2=3m^2$
($\tau=\frac{2\pi}{3}$).
In the special cases $p^2=m^2$ and $p^2=2m^2$ we get,
respectively,
\bea
J_3(1,1,1)\bigr|_{p_i^2=m^2, \; n=4}
&=& \frac{{\rm i} \pi^2}{m^2 \sqrt{3}}
\Biggl[ \Cl{2}{\frac{\pi}{3}}-\pi\ln{2} \Biggl] \; ,
\\
J_3(1,1,1)\bigr|_{p_i^2=2m^2, \; n=4}
&=& \frac{{\rm i} \pi^2}{4 m^2 \sqrt{3}}
\Biggl[ 5 \Cl{2}{\frac{\pi}{3}} 
- 3\pi \ln\left(2+\sqrt{3}\right)
\Biggl] \; .
\eea

To analytically continue the result~(\ref{DD(5.17)sym}) to
other regions of interest, we can use the representation
(\ref{DD(5.16)sym}). We note that $\tan^2\eta<0$ for $p^2>3m^2$
and for $p^2<0$. To describe the Euclidean region ($p^2=-\mu^2<0$),
as well as the  
region above the two-particle threshold ($p^2>4m^2$),
it is convenient to introduce the angle $\tau'$ such that
\be
\cos\tau'=\frac{6m^2-p^2}{2(3m^2-p^2)} 
=\frac{6m^2+\mu^2}{2(3m^2+\mu^2)} \; .
\ee
The argument of the logarithm in Eq.~(\ref{DD(5.16)sym}) is always 
positive for $p^2<0$,
and always negative for $3m^2<p^2<4m^2$. In the
region $p^2>4m^2$,
the argument is positive for $0<\phi<\frac{1}{2}(\pi-\tau')$ and
negative for $\frac{1}{2}(\pi-\tau')<\phi<\frac{\pi}{3}$.
This means that we obtain an imaginary part for $3m^2<p^2<4m^2$
and for $p^2>4m^2$. 

The occurring angular integrals can be calculated by
using Eqs.~(33), (34), (36) and (38) on p.~308 of~\cite{Lewin}.
In this way, we arrive at the following results.
For $3m^2<p^2<4m^2$, we have
\bea
\label{sym34}
J_3(1,1,1)\bigr|_{p_i^2=p^2, \; n=4}
&=& -\frac{{\rm i} \pi^2 \sqrt{3}}{p^2}
\Biggl[ \Cl{2}{\frac{2\pi}{3}+\tau}
+\Cl{2}{\frac{2\pi}{3}-\tau}-2\Cl{2}{\frac{2\pi}{3}}
\nonumber \\ &&
+ \tau \ln\left(
\frac{\sin\left(\frac{\tau}{2}+\frac{\pi}{3}\right)}
     {\sin\left(\frac{\tau}{2}-\frac{\pi}{3}\right)} \right)
+ \frac{2 {\rm i} \pi^2}{3}
\Biggl] .
\eea
For $p^2>4m^2$, we have
\be
\label{sym>4}
J_3(1,1,1)\bigr|_{p_i^2=p^2, \; n=4}
= -\frac{{\rm i} \pi^2 \sqrt{3}}{p^2}
\Biggl[ \Cl{2}{\frac{\pi}{3}\!+\!\tau'}
+\Cl{2}{\frac{\pi}{3}\!-\!\tau'}-2\Cl{2}{\frac{\pi}{3}}
+ {\rm i} \pi \left( \tau'-\frac{\pi}{3}\right)
\Biggl] .
\ee
In particular, at $p^2=4m^2$ we get
\be
J_3(1,1,1)\bigr|_{p_i^2=4m^2, \; n=4}
= \frac{{\rm i} \pi^2}{2m^2\sqrt{3}}
\left[ 5 \Cl{2}{\frac{\pi}{3}} - {\rm i}\pi^2 \right] \; .
\ee
Finally, in the Euclidean region, $p^2=-\mu^2<0$,
we get
\be 
\label{sym<0}
J_3(1,1,1)\bigr|_{p_i^2=-\mu^2, \; n=4}
= \frac{{\rm i} \pi^2 \sqrt{3}}{\mu^2}
\Biggl[ \Cl{2}{\frac{\pi}{3}+\tau'}
+\Cl{2}{\frac{\pi}{3}-\tau'}-2\Cl{2}{\frac{\pi}{3}}
\Biggl] \; .
\ee
The latter result gives the correct massless limit~\cite{CelGon},
\be
J_3(1,1,1)\bigr|_{p_i^2=-\mu^2,\; m=0, \; n=4}
= -\frac{4{\rm i}\pi^2}{\mu^2\sqrt{3}} \Cl{2}{\frac{\pi}{3}}.
\ee

In Fig.~2 we show the function $\varphi_3(p^2,p^2,p^2)$, which in
$n=4$ dimensions is given by $J_3(1,1,1)/({\rm i}\pi^2)$.
The real part is seen to be singular at the anomalous threshold,
$p^2=3m^2$, whereas it has a kink (the first derivative
is discontinuous) at the normal (two-particle) threshold, $p^2=4m^2$.
The imaginary part starts at a finite value for $p^2=3m^2$,
and has a kink at $p^2=4m^2$.
\begin{figure}[htb]
\refstepcounter{figure}
\label{Fig:phi3}
\addtocounter{figure}{-1}
\begin{center}
\setlength{\unitlength}{1cm}
\begin{picture}(13.0,9.0)
\put(-1,0){
\mbox{\epsfysize=10cm\epsffile{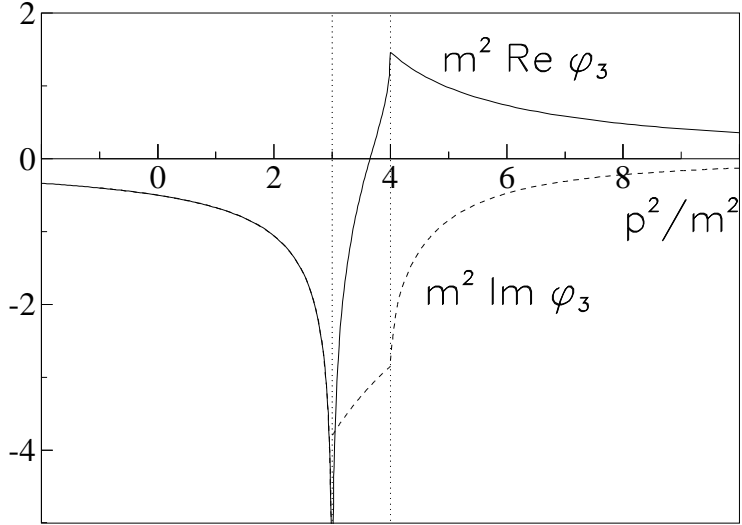}}}
\end{picture}
\vspace*{-4mm}
\caption{The integral $\varphi_3(p^2,p^2,p^2)$ 
in four dimensions, $n=4$.}
\end{center}
\end{figure}

\subsection{On-shell and zero-momentum cases}

When all momenta squared vanish, we get a tadpole,
\be
J_3(1,1,1)\bigl|_{p_1^2=p_2^2=p_3^2=0}
= -\frac{1}{2} {\rm i}\pi^{n/2} (m^2)^{n/2-3} \Gamma(3-n/2) \; .
\ee

When two momenta are on shell, $p_1^2=p_2^2=0$,
the integral $J_3(1,1,1)$ can be presented 
in arbitrary dimension as (see, e.g., Eq.~(40)
of~\cite{BD-TMF}):
\be
\label{two_os}
J_3(1,1,1)\bigl|_{p_1^2=p_2^2=0}
= -\frac{1}{2} {\rm i}\pi^{n/2} (m^2)^{n/2-3} \Gamma(3-n/2)\;
_3F_2\biggl( \begin{array}{c}
3-n/2, \; 1, \; 1 \\ 3/2, \; 2 \end{array}
\biggl|~\frac{p_3^2}{4m^2} \biggl) \; ,
\ee
where $_PF_Q$ is the generalized hypergeometric function.
A number of integral representations for this integral
are given in Sec.~3.3 of~\cite{DK}, where also explicit
results for some terms of the $\varepsilon$-expansion 
are presented. In four dimensions,
\bea
\label{elementary}
_3F_2\Biggl( \begin{array}{c} 1,\; 1,\; 1 \\ 3/2,\; 2 \end{array}
\biggl|~z\Biggl) =
\Biggl\{ \begin{array}{l}
z^{-1}\arcsin^2\sqrt{z} \; , \hspace{24mm} z\geq 0\; , \\
-z^{-1} \ln^2\left(\sqrt{1-z}+\sqrt{-z}\right), \quad z\leq 0\; .
\end{array} 
\eea
This is the familiar result~\cite{H->2gamma} for the case 
of a Higgs particle coupling
to two massless vector particles, $Hgg$ or $H\gamma\gamma$
(see also in~\cite{HHG}).

In the case when just one momentum is on shell, $p_3^2=0$,
we can start from the general hypergeometric representation
(see Eq.~(37) of~\cite{BD-TMF}). Putting $p_3^2=0$
we get a double hypergeometric series. Then, denoting
$j_1+j_2=j$ and summing over the remaining index, we
arrive at
\bea
\label{one_os}
J_3(1,1,1)\bigl|_{p_3^2=0}
&\!\!=\!\!& -\frac{1}{2} {\rm i}\pi^{n/2} 
(m^2)^{n/2-3} \Gamma(3-n/2)\;
\frac{1}{p_1^2-p_2^2}
\nonumber \\
&& \times \Biggl[
p_1^2\; _3F_2\biggl( \begin{array}{c}
3-n/2, 1, 1 \\ 3/2, \; 2 \end{array}
\biggl|~\frac{p_1^2}{4m^2} \biggl)
- p_2^2\; _3F_2\biggl( \begin{array}{c}
3-n/2, 1, 1 \\ 3/2, \; 2 \end{array}
\biggl|~\frac{p_2^2}{4m^2} \biggl)
\Biggl] \; ,
\hspace*{6mm}
\eea
or
\be
\label{one_os_}
J_3(1,1,1)\bigl|_{p_3^2=0}
= \frac{1}{p_1^2-p_2^2}
\biggl[ p_1^2\; J_3(1,1,1)\bigl|_{p_2^2=p_3^2=0}
\; - \; p_2^2\; J_3(1,1,1)\bigl|_{p_1^2=p_3^2=0}
\biggr] \; ,
\ee
i.e., it is just a linear combination of the 
integrals~(\ref{two_os}) with two legs on shell. 
In particular, Eq.~(\ref{one_os_}) means that all results of 
Sec.~3.3 of~\cite{DK} are applicable to the case of
two legs off shell, too.
For $n=4$, Eq.~(\ref{one_os_}) yields a combination of elementary 
functions~(\ref{elementary}) --- this result is known
from the calculation of the one-loop 
$H \to Z\gamma$ vertex~\cite{H->Zgamma} (see also in~\cite{HHG}).

In the case when $p_3=0$ ($p_1=-p_2\equiv p$),
the three-point integral $J_3$ effectively becomes 
a two-point integral, with one of the massive 
denominators to the second power. Formally, we can write
\be
J_3(1,1,1)\bigl|_{p_3=0} = J_3(0,2,1) \; .
\ee
Using integration by parts~\cite{ibp} (see also Eq.~(A.17)
of~\cite{BDS}), $J_3(0,2,1)$ can be reduced to $J_3(0,1,1)$
and a tadpole integral. In this way, we obtain
\be
J_3(1,1,1)\bigl|_{p_3=0}
= \frac{1}{2(4m^2-p^2)}
\biggl[ 2(n-3) J_3(0,1,1) - \frac{1}{m^2} (n-2) J_3(0,0,1)
\biggr] \; ,
\ee
or, in the notation of Eqs.~(\ref{J_3 defs})--(\ref{kt def}),
\be
\varphi_3(p^2,p^2,0) = \frac{1}{2(4m^2-p^2)}
\bigl[ 2(n-3) \kappa_2(p^2) 
- (n-2) {\widetilde{\kappa}} \bigr] \; .
\ee

\section{Conclusions}

We have obtained general results for the quark-loop contributions 
to the three-gluon vertex in an arbitrary dimension.
For the general off-shell case, the decomposition of Ball and
Chiu~\cite{BC2} was used. 
The general off-shell case, as well as all
on-shell limits of interest have been considered. 
The obtained results satisfy the corresponding Ward--Slavnov--Taylor 
identity~(\ref{ggg_WST_one-loop_massive}).
We have also presented some results for the corresponding
three-point integrals. 

This calculation completes the investigation of the one-loop
three-gluon vertex in an arbitrary dimension, which was
initiated in~\cite{DOT1}. 
This could be a valuable element in two-loop (and higher) calculations.
Moreover, together with Ref.~\cite{DOS} the calculation 
completes the study of one-loop three-point vertices in QCD, in an
arbitrary covariant gauge.

A similar study of the two-loop-order corrections to the
three-gluon vertex (and other QCD vertices) requires more involved 
techniques.
We note that for some special configurations (for massless quarks) 
two-loop results for the three-gluon vertex are already available,
in particular, for the 
zero-momentum case~\cite{BL,DOT2} and the $p_1^2=p_2^2=0$
case~\cite{DO1}. A numerical approach to the symmetric case 
has been developed in~\cite{ChR}. 
Moreover, the three-loop 
results in the zero-momentum case are available in~\cite{ChS}.

\vspace{3mm}

{\bf Acknowledgements.}
It is a pleasure to thank Oleg Tarasov for constructive comments.
A.~D. is grateful to M.Yu.~Kalmykov for useful discussions.
This research has been supported by the Research Council of Norway.
L.~S.'s research was supported by the Norwegian State Educational 
Loan Fund.
A.~D.'s research was supported by the Deutsche Forschungsgemeinschaft.
Partial support from the RFBR grant 01-02-16171 is also acknowledged.



\begin{thebibliography}{99}

\bibitem{QCD} 
H.~Fritzsch, M.~Gell-Mann and H.~Leutwyler,
         {\em Phys.\ Lett.} 47B (1973) 365; \\
D.J.~Gross and F.~Wilczek, {\em Phys.\ Rev.\ Lett.} 30 (1973) 1343; \\
H.D.~Politzer, {\em Phys.\ Rev.\ Lett.} 30 (1973) 1346; \\
S.~Weinberg, {\em Phys.\ Rev.\ Lett.} 31 (1973) 494.

\bibitem{QCD-reviews}
CTEQ Collaboration, G. Sterman {\it et al.},
{\em Rev.~Mod.~Phys.} {\bf 67} (1995) 157; \\
Z.~Bern, L.~Dixon and D.A.~Kosower,
{\em Ann.~Rev.~Nucl.~Part.~Sci.} {\bf 46} (1996) 109,\\
S.~Catani {\it et al.},
hep-ph/0005025 and
hep-ph/0005114.

\bibitem{CelGon}
W.~Celmaster and R.J.~Gonsalves,
{\em Phys.\ Rev.} {\bf D20} (1979) 1420.

\bibitem{PT}
P.~Pascual and R.~Tarrach, {\em Nucl.\ Phys.} {\bf B174} (1980) 123.

\bibitem{NPS}
M.A.~Nowak, M.~Prasza{\l}owicz and W.~S{\l}omi{\'n}ski,
{\em Ann.\ Phys.\ (N.Y.)} {\bf 166} (1986) 443.

\bibitem{BC2}
J.S.~Ball and T.-W.~Chiu,
 {\em Phys.\ Rev.} {\bf D22} (1980) 2550;
{\bf D23} (1981) 3085 (E).

\bibitem{BF}
F.T.~Brandt and J.~Frenkel, {\em Phys.\ Rev.} {\bf D33} (1986) 464.

\bibitem{DOT1}
A.I.~Davydychev, P.~Osland and O.V.~Tarasov,
{\em Phys.\ Rev.} {\bf D54} (1996) 4087; {\bf D59} (1999) 109901 (E).

\bibitem{DOT-Rh}
A.I.~Davydychev, P.~Osland and O.V.~Tarasov,
{\em Nucl. Phys. B (Proc. Suppl.)} {\bf 51C} (1996) 289.

\bibitem{DOS}
A.I.~Davydychev, P.~Osland and L.~Saks,
{\em Phys.\ Rev.} {\bf D63} (2001) 014022.

\bibitem{dimreg}
G.~'t~Hooft and M.~Veltman, {\em Nucl.\ Phys.} {\bf B44} (1972) 189;\\
C.G.~Bollini and J.J.~Giambiagi, {\em Nuovo Cim.} {\bf 12B} (1972) 20;\\
J.F.~Ashmore, {\em Lett.~Nuovo~Cim.} {\bf 4} (1972) 289;\\
G.M.~Cicuta and E.~Montaldi, {\em Lett.~Nuovo~Cim.} {\bf 4} (1972) 329.

\bibitem{AA}
E.~Abdalla and M.C.B.~Abdalla,
{\em Phys.\ Rep.} {\bf 265} (1996) 253. 

\bibitem{lattice}
C.~Parrinello, {\em Phys.\ Rev.} {\bf D50} (1994) 4247;\\
D.~Becirevic {\em et al.},
{\em Nucl.\ Phys. B (Proc.\ Suppl.)} {\bf 83} (2000) 159 
(hep-lat/9908056); \\
P.~Boucaud {\em et al.}, 
{\em Phys.\ Lett.} {\bf B493} (2000) 315. 

\bibitem{DD} 
A.I.~Davydychev and R.~Delbourgo,
{\em J.\ Math.\ Phys.} {\bf 39} (1998) 4299.

\bibitem{Kn}
B.A.~Kniehl, 
{\em Phys.\ Rep.} {\bf 240} (1994) 211. 

\bibitem{HHG}
J.F.~Gunion, H.E.~Haber, G.~Kane and S.~Dawson,
{\em The Higgs Hunter's Guide} (Addison-Wesley, Reading, 1990).

\bibitem{KB}
S.K.~Kim and M.~Baker, {\em Nucl.\ Phys.} {\bf B164} (1980) 152.

\bibitem{BD-TMF} E.E.~Boos and A.I.~Davydychev,
{\em Teor.\ Mat.\ Fiz.} {\bf 89} (1991) 56
[{\em Theor.\ Math.\ Phys.} {\bf 89} (1991) 1052].

\bibitem{MarPag}
W.~Marciano and H.~Pagels, 
{\em Phys.\ Rep.} {\bf 36} (1978) 137.

\bibitem{PT-QCD}
P.~Pascual and R.~Tarrach, {\em QCD: Renormalization
     for the Practitioner}, Lecture Notes in Physics, Vol.~194
 (Springer, Berlin, 1984).

\bibitem{reduce}
A.C.~Hearn,  {\em REDUCE User's Manual} (version 3.6),
RAND publication CP78  (Santa Monica, 1995).

\bibitem{BL} 
E.~Braaten and J.P.~Leveille, 
{\em Phys.\ Rev.} {\bf D24} (1981) 1369.

\bibitem{DOT2}
A.I.~Davydychev, P.~Osland and O.V.~Tarasov,
{\em Phys.\ Rev.} {\bf D58} (1998) 036007.

\bibitem{jmp1}
A.I.~Davydychev,  {\em J.~Math.~Phys.} {\bf 32} (1991) 1052.

\bibitem{JPA}
A.I.~Davydychev, {\em J. Phys.} {\bf A25} (1992) 5587.

\bibitem{ibp} 
F.V.~Tkachov, {\em Phys.\ Lett.} {\bf 100B} (1981) 65; \\
K.G.~Chetyrkin and F.V.~Tkachov, {\em Nucl.\ Phys.} {\bf B192}
          (1981) 159.

\bibitem{Nickel}
B.G.~Nickel, {\em J.~Math.~Phys.} {\bf 19} (1978) 542.

\bibitem{tHV'79}
G.~'t Hooft and M.~Veltman,
{\em Nucl.\ Phys.} {\bf B153} (1979) 365.

\bibitem{vOV}
G.J. van Oldenborgh and J.A.M.~Vermaseren,
{\em Z.\ Phys.} {\bf C46} (1990) 425.

\bibitem{Lewin} L.~Lewin, {\em Polylogarithms and Associated
Functions} (North-Holland, Amsterdam, 1981).

\bibitem{Wagner}
N.~Ortner and P.~Wagner,
{\em Ann.~Inst.~Henri~Poincar\'e (Phys.~th\'eor.)}   
 {\bf 63} (1995) 81;\\
P.~Wagner, {\em Indag.~Math.} {\bf 7} (1996) 527.

\bibitem{Crete}
A.I.~Davydychev,
Proc.\ Workshop ``AIHENP-99'', Heraklion, Greece, April 1999 (Parisianou S.A.,
Athens, 2000), p.~219
(hep-th/9908032).

\bibitem{DK} A.I.~Davydychev and M.Yu.~Kalmykov,
hep-th/0012189 (to appear in {\em Nucl.\ Phys.} {\bf B}).

\bibitem{bastei_tar}
O.V.~Tarasov, 
{\em Nucl. Phys. B (Proc.\ Suppl.)} {\bf 89} (2000) 237
(hep-ph/0102271).

\bibitem{singularities}
L.D.~Landau, {\em Nucl.\ Phys.} {\bf 13} (1959) 181;\\
G.~K\"all\'en and A.~Wightman,
{\em Mat.\ Fys.\ Skr.\ Dan.\ Vid.\ Selsk.} {\bf 1} (No.~6) (1958) 1;\\
R.~Blankenbecler and Y.~Nambu, {\em Nuovo Cim.} 
{\bf 18} (1960) 580.

\bibitem{H->2gamma}
A.I.~Vainshtein, M.B.~Voloshin, V.I.~Zakharov and M.A.~Shifman,
{\em Yad.\ Fiz.} {\bf 30} (1979) 1368
[{\em Sov.\ J.\ Nucl.\ Phys.} {\bf 30} (1979) 711].

\bibitem{H->Zgamma}
L.~Bergstr\"om and G.~Hulth, 
{\em Nucl.\ Phys.} {\bf B259} (1985) 137;\\
J.F.~Gunion, G.L.~Kane and J.~Wudka, 
{\em Nucl.\ Phys.} {\bf B299} (1988) 231.

\bibitem{BDS}
F.A.~Berends, A.I.~Davydychev and V.A.~Smirnov,
{\em Nucl.\ Phys.} {\bf B478} (1996) 59.  

\bibitem{DO1}
A.I.~Davydychev and P.~Osland, 
{\em Phys. Rev.} {\bf D59} (1999) 014006.

\bibitem{ChR}   
K.G.~Chetyrkin and A.~R\'etey, hep-ph/0007088.

\bibitem{ChS}
K.G.~Chetyrkin and T.~Seidensticker, 
{\em Phys.\ Lett.} {\bf B495} (2000) 74.

\end{thebibliography}
\end{document}